\documentclass{jfm}

\usepackage{graphicx}
\usepackage{amsmath}
\usepackage{mathtools}
\usepackage{newtxtext}
\usepackage{newtxmath}
\usepackage{natbib}
\usepackage{tensor}
\usepackage{amssymb}
\usepackage{hyperref}
\hypersetup{
    colorlinks = true,
    urlcolor   = blue,
    citecolor  = black,
}

\newcommand{\RomanNumeralCaps}[1]

\newcommand{\bx}{\boldsymbol{x}}
\newcommand{\bff}{\boldsymbol{f}}
\newcommand{\bX}{\boldsymbol{X}}

\newcommand{\bY}{\boldsymbol{Y}}
\newcommand{\bu}{\boldsymbol{u}}

\newcommand{\xii}{\boldsymbol{\mathrm{\xi}}}
\newcommand{\Xii}{\boldsymbol{\mathrm{\Xi}}}

\newcommand{\Lbar}[1]{\overline{#1}^{ \mathrm{\hspace{0.5mm}\small L}} }
\newcommand{\Stk}[1]{\overline{#1}^{ \mathrm{\hspace{0.5mm}\small S}} }\def\e{\mathrm{e}}

\def\e{\mathrm{e}}

\title{Grid-based calculation of Lagrangian mean}

\author{Hossein A. Kafiabad\aff{1}\corresp{\email{h.kafiabad@ed.ac.uk}} }

\affiliation{\aff{1}School of Mathematics and Maxwell Institute for Mathematical Sciences, \\
University of Edinburgh, Edinburgh, UK}

\begin{document}
\maketitle

\begin{abstract}
Lagrangian averaging has been shown to be more effective than Eulerian mean in separating waves from slow dynamics in two-timescale flows. It also appears in many reduced models that capture the wave feedback on the slow flow. Its calculation, however, requires tracking particles in time, which imposes several difficulties in grid-based numerical simulations or estimation from fixed-point measurements. To circumvent these difficulties, we propose a grid-based iterative method to calculate Lagrangian mean without tracking particles in time, which also reduces computation, memory footprint and communication between processors in parallelised numerical models. To assess the accuracy of this method several examples are examined and discussed. We also explore an application of this method in the context of shallow-water equations by quantifying the validity of wave-averaged geostrophic balance -- a modified form of geostrophic balance accounting for the effect of strong waves on slow dynamics.
\end{abstract}

\section{Introduction}\label{sec:intro}

Large-scale geophysical flows are characterised by fast and slow timescales; the fast dynamics consists of inertia-gravity waves, and the slow dynamics is usually described by geostrophic and hydrostatic balance \citep{vanneste2013balance,vallis2017atmospheric}. There is a particular interest in decomposing the flow into these components and analysing them separately. Focusing on the slow dynamics reduces the governing equations to a simpler set that is easier to interpret and solve \citep{charney1971geostrophic,warn1997nonlinear,mcintyre2000potential,mohebalhojeh2001hierarchies}. Likewise, mathematical tools specific to wave dynamics may be employed to analyse the fast component \citep{buhler2009waves}. In addition to bringing physical insights, such decompositions are used to parametrise the small-scale dynamics that are not resolved in ocean and weather models \citep{sutherland2019recent}.

The most straightforward decomposition is averaging the variables over the fast timescale at fixed spatial points to derive the \emph{Eulerian} mean. Although easy to implement, this is not the most effective way of extracting wave signal from the mean-flow.  For instance, the Doppler shifting of the wave frequency by strong mean-flows can eclipse the inherent timescale separation between them. To circumvent this issue, one can derive the \emph{Lagrangian} mean (LM) instead by averaging the flow variables along particle trajectories and assigning the mean to a fixed particle label. Recent studies have shown a range of oceanic phenomena are better understood with Lagrangian averaging, including the spontaneous wave generation \citep{nagai2015spontaneous,shakespeare2017spontaneous,shakespeare2018life}, internal tides \citep{shakespeare2019momentum} and their effect on the ecology of the coastal and reef zones \citep{bachman2020particle}.

Another motivation for calculating the LM comes from its natural appearance in reduced models that are derived by wave averaging \citep{bretherton1971general,grimshaw1975nonlinear,andrews1978exact,buhler2009waves,gilbert2018geometric}. A striking prediction, highlighting the fundamental role of the LM, is that the hydrostatic and geostrophic balance continues to hold in the presence of strong waves for LM quantities \citep[][and references therein]{kafiabad2021wave}.Similarly, models of near-inertial waves interacting with the mean-flow naturally involve LM variables \citep{xie2015generalised,wagner2015available,wagner2016three,asselin2020penetration}. In general, these studies show that the slow dynamics with the inclusion of wave feedback on the flow is more accurately represented by LM quantities. To examine the validity of these models and utilise them in operational and research models there is a growing need to compute the LM. 

Computing the LM, however, requires more effort than computing the Eulerian mean. Most numerical models are discretised using spatial grid points and the majority of measurements are taken at fixed locations in space. To derive the LM, in these cases passive particles are seeded in the flow and advected (forward or backward in time) based on the interpolated velocities at particle positions \citep[e.g.][]{nagai2015spontaneous,shakespeare2017spontaneous,shakespeare2018life,shakespeare2019momentum}. This requires saving the time series of several fields, which are not normally saved in numerical models, at high spatial resolution. The other issue with particle tracking comes up in parallelised solvers, where the computational domain is divided into many subdomains assigned to independent processors. Even uniformly seeded particles may leave their initial subdomain, causing additional communication between the processors and disturbing the computational load balance among them.

We propose a grid-based Lagrangian averaging (GBLA) method that does not require particle tracking. Instead, the LM is updated on grid points using the mean associated with the trajectories that pass through the grid points in the previous timestep. As a result, there is no need for saving any additional time series. This method can be integrated with numerical models to efficiently calculate the LM on the fly. It also maintains a load balance between the processors in parallel implementations and minimises the communication between them. We first describe this method and its mathematical justification in \S \ref{sec:method_description}. We then investigate its validity in five examples in  \S \ref{sec:validationAll}. An application of GBLA is considered in \S \ref{sec:application}, where we study the wave-averaged geostrophy for the rotating shallow-water equations. The paper concludes with a discussion in \S \ref{sec:discussion}.

\section{Grid-based Lagrangian averaging}\label{sec:method_description}

\subsection{Algorithm}

Let us consider the LM of the field $\phi(\bx,t)$ from $t_0$ to $t_0+T$. Before explaining the steps of GBLA, we clarify some notations and definitions. We denote the Eulerian-mean of $\phi(\bx,t)$ over the period of $t_0$ to $t$ with
\begin{equation}\label{EuMean}
\overline{\phi}(\bx,t_0,t) = \frac{1}{t-t_0} \int_{t_0}^{t} \phi(\bx,\tau) d \tau.
\end{equation}
%To explain our method easier we assign the Eulerian-mean to the end-point of averaging interval. It may be preferred to choose the middle-point instead \citep[e.g.][]{wagner2015available}, which can be obtained by a simple shift in time  $t \to t-T/2$.
The LM of $\phi$ is the time-average of this quantity along the trajectory, that is, at fixed particle label instead of fixed spatial position. There are different ways of labelling particles; for instance, the particle position at the beginning or the end of averaging interval may be selected as labels. Denoting this label by $\bx$, we can map it to the particle position at time $t$ (denoted by $\bX$) using the particle displacement field $\xii(\bx,t)$,
\begin{equation}\label{mapping} 
  \bX = \Xii(\bx,t) = \bx+\xii(\bx,t).
\end{equation}   
Using \eqref{EuMean} and \eqref{mapping}, the LM operator is mathematically described as 
\begin{equation}\label{LagMean}
\Lbar{\phi}(\bx,t_0,t) = \frac{1}{t-t_0} \int_{t_0}^{t} \phi(\bx+\xii(\bx,\tau),\tau) d \tau. 
\end{equation}
Different choices of $\bx$ lead to different displacement fields $\xii$. \cite{andrews1978exact} defined the `generalised Lagrangian-mean (GLM)' by requiring
\begin{equation}\label{GLMdisplacement}
\overline{\xii}(\bx,t) = 0.
\end{equation}
In other words, $\bx$ is now the mean position of the particle from $t_0$ to $t$. If the displacement field is small (compared to the changes in mean position), \eqref{GLMdisplacement} gives the LM some Eulerian-like character as $\bx$ is more than just a label and represents the spatial location of the trajectory for which the mean is calculated.  As it will be shown in the examples of \S \ref{sec:1Dexamples}, this choice of label is more effective at removing the fast timescale in the LM estimate provided that the mapping  \eqref{mapping} does not have a fast time dependence. If other labels such as instantaneous position at the end or middle of averaging interval are chosen, the filtered signal may still contain some fast oscillations or get distorted  (see the numerical example of \S \ref{sec:1Dexamples}). Moreover, the GLM definition holds properties that lead to significant analytical simplifications in modelling two timescale geophysical flows \citep{andrews1978exact,buhler1998non-dissipative,xie2015generalised,gilbert2018geometric}. Throughout this paper, we use  \eqref{LagMean} in conjunction with \eqref{GLMdisplacement} to express the LM in terms of the mean position. For $t<(t_0+T)$, \eqref{EuMean} and \eqref{LagMean} represent \textit{partial means}, and for $t = t_0+T$ the complete mean for the intended averaging interval. Considering that in this section we focus only on one averaging interval with a fixed starting point, we drop $t_0$ in the arguments of the mean operators to lighten the notation.

The core idea behind GBLA is to calculate partial LMs recursively in time. For instance, assume that the partial LM from $t_0$ to $t_k$ ($t_0<t_k<t_0+T$) for the particles that reach the grid points at time $t_k$ is already computed. Through the steps laid out below, the LM is extended to $t_{k+1}$ and calculated for a new set of particles that reach the grid points at time $t_{k+1}$ (using the partial averages of the previous step). In this approach, at each timestep a different set of trajectories is considered (the set that ends up at grid points at that timestep), which is in contrast with particle tracking methods where one trajectory is followed all along over the entire averaging interval. The incremental extension of partial LMs continues to the end of averaging interval $t_0+T$. The completed LM field is then assigned to time $t_0+T/2$. Note that in our demonstration we focus only on one averaging interval, but this process can be repeated for new intervals, and averaging procedure can be carried out simultaneously. For example, the next interval may start from $t_m$ and end in $t_{m+N}$ and so on. If the beginning of a later interval exceeds $t_0+T$, the memory allocated to the first interval (from $t_0$ to $t_0+T$) can be freed in a systematic way and used for the new intervals.

\begin{figure}
    \centering
     \includegraphics[width=.55\linewidth]{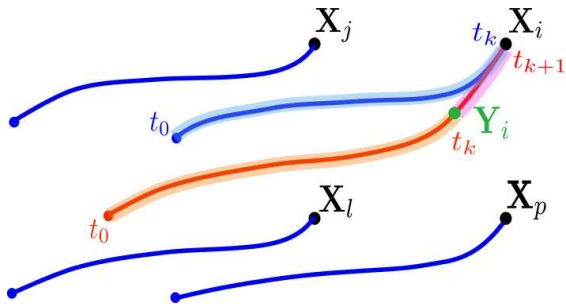}
\caption{Schematic view of particle trajectories that pass through the grid points $\bX_i$, $\bX_j$, $\bX_p$ and $\bX_l$ that are marked by black. The trajectories that reach these grid points at $t_k$ are marked by blue and the trajectory that reaches the grid point at $t_{k+1}$ by red. For $\bX_i$, the times associated with the particle position of red and blue trajectory are shown as well. }
\label{fig:schem_traj}
\end{figure}

Equipped with the above definitions, we present the steps to compute $\Lbar{\phi}(\Xii^{-1}(\bX_i,t_{k+1}),t_{k+1})$ from $\Lbar{\phi}(\Xii^{-1}(\{\bX_{\alpha}\},t_k),t_k)$, where $\bX_i$ is an arbitrary grid point in space and $\{\bX_{\alpha}\}$ is a set consisting of $\bX_i$ and its neighbours. In other words, our steps show how to calculate the partial LM (from $t_0$ to $t_{k+1}$) for the particle that passes through $\bX_i$ at $t_{k+1}$  by knowing the LM (from $t_0$ to $t_k$) for the particles that pass through $\bX_i$ and its neighbouring grid points at $t_{k}$. To simplify the notation we use one spatial index $i$, which can be viewed as a multi-element vector or replaced by several indices in higher than one dimension. Considering that $\Lbar{\phi}(\Xii^{-1}(\bX_i,t_k),t_k)$ is the partial LM for the particle that reaches $X_i$ at $t_{k}$ and $\Lbar{\phi}(\Xii^{-1}(\bX_i,t_{k+1}),t_{k+1})$ is the partial LM for the particle that reaches $X_i$ at $t_{k+1}$, these two LMs belong to two separate trajectories (marked by red and blue in figure  \ref{fig:schem_traj}).

\begin{itemize}
\item[\textbf{Step 1}] Calculate $\phi(\bX_i, t_{k+1})$ and the velocity $\bu(\bX_i, t_{k+1})$ by integrating the governing equations from $t_k$ to $t_{k+1}$. Note that a finer timestep may be used for the time integration of the governing equations than that used for averaging; there might be several simulation timesteps between $t_k$ and $t_{k+1}$.

\item[\textbf{Step 2}] Consider the particle that goes through $\bX_i$ at time $t_{k+1}$ (red trajectory in figure  \ref{fig:schem_traj}). Approximate the position of this particle at time $t_k$ by advecting it backward in time
\begin{equation}\label{advect_backward}
	\bY_i = \bX_i - (t_{k+1}-t_k) \bu(\bX_i, t_{k+1}) .
\end{equation}
This point is marked by green in figure \ref{fig:schem_traj} and splits the red trajectory into two parts.  
\item[\textbf{Step 3}] Denote the partial LM of $\phi$ between $t_0$ and $t_k$ for the red trajectory by $\Lbar{\phi}(\Xii^{-1}(\bY_i,t_k),t_k)$. Then, find its value by interpolation using the LM for the trajectories that go through the neighbouring grid points (marked by blue curves). For example, $\bY_i $ lies between $\{\bX_{\alpha}\}$ for ${\alpha} = i, j, l$ and $p$ in figure \ref{fig:schem_traj}. Therefore, the four values of $\Lbar{\phi}(\Xii^{-1}(\{\bX_{\alpha}\},t_k),t_k)$ may be used to interpolate $\Lbar{\phi}(\Xii^{-1}(\bY_i,t_k),t_k)$. Note that $\Lbar{\phi}(\Xii^{-1}(\bY_i,t_k),t_k)$ and $\Lbar{\phi}(\Xii^{-1}(\bX_i,t_k),t_k)$ are both averages over $[t_0,\ t_k]$ but for different trajectories (highlighted by orange and light blue respectively). %The trajectory associated with $\Lbar{\Phi_i}(n)$  passes through $\bX_i $ at timestep $n+1$, and the one associated with $\Lbar{\phi_i}(n)$  passes through $\bX_i $ at timestep $n$.
\item[\textbf{Step 4}] Selecting the mean position of the red trajectory from $t_0$ to $t_{k+1}$ as a label for the particle that passes through $\bX_i$ at $t_{k+1}$ and denoting it by  $\bx_i$, use the definition \eqref{LagMean} to compute
\begin{align}\label{updateLM}
\begin{split}
\Lbar{\phi}( & \Xii^{-1}(\bX_i,t_{k+1}),t_{k+1}) =  
\frac{1}{t_{k+1}-t_0} \int_{t_0}^{t_{k+1}} \phi(\bx_i + \xii(\bx_i,\eta),\eta) d \eta \\
& = \frac{t_k-t_0}{t_{k+1} - t_0} \ \frac{1}{t_{k} - t_0}  \int_{t_0}^{t_k} \phi(\bx_i + \xii(\bx_i,\eta),\eta) d \eta+ \frac{1}{t_{k+1} - t_0} \int_{t_k}^{t_{k+1}} \phi(\bx_i + \xii(\bx_i,\eta),\eta) d \eta \\
& = \frac{t_k-t_0}{t_{k+1} - t_0}\ \Lbar{\phi}(\Xii^{-1}(\bY_i,t_k),t_k) +\frac{1}{t_{k+1} - t_0} \int_{t_k}^{t_{k+1}} \phi(\bx_i + \xii(\bx_i,\eta),\eta) d \eta.
\end{split}
\end{align}
The first term in RHS is already computed in the previous step. The second term, which is the weighted mean of the red trajectory from $t_k$ to $t_{k+1}$, can be approximated by 
\begin{equation}\label{approx1}
  \int_{t_k}^{t_{k+1}} \phi(\bx_i + \xii(\bx_i,\eta),\eta) d \eta \approx \phi(\bX_i,t_{k+1})\ (t_{k+1}-t_k).
\end{equation}
A better approximation may be derived by the trapezoidal rule
\begin{equation}\label{approx2}
 \int_{t_k}^{t_{k+1}} \phi(\bx_i + \xii(\bx_i,\eta),\eta) d \eta \approx \frac{1}{2}\left[   \phi(\bY_i,t_{k}) +\phi(\bX_i,t_{k+1}) \right] (t_{k+1}-t_k).
\end{equation}
To calculate $\phi(\bY_i,t_{k})$ another interpolation is required. Hence, depending on the desired accuracy and computational cost, \eqref{approx2} may or may not be preferred to \eqref{approx1}. 

So far the calculation of the LM field in \eqref{updateLM} is complete, but it is given for the end position of each particle $\bX_i$. If the definition of GLM with the condition \eqref{GLMdisplacement} is desired, the LM should be assigned to the mean position of particles. In other words, we need to calculate $\bx_i = \Xii^{-1}(\bX_i,t_{k+1})$, which maps the particle's end position $\bX_i$ to its mean from $t_0$ to $t_{k+1}$. To calculate the mean position, we update it in a similar manner to \eqref{updateLM} at each timestep via step 5.
\item[\textbf{Step 5}]  Assuming $\Xii^{-1}(\bX,t_{k}) $ is known for all grid points from the previous timestep, compute the mean position of the red trajectory from $t_0$ to $t_{k+1}$
\begin{align}\label{updateMeanPos}
\begin{split}
	  \Xii^{-1}(\bX_i,t_{k+1}) &= \frac{1}{t_{k+1}-t_0} \int_{t_0}^{t_{k+1}} \bx_i + \xii(\bx_i,\eta) d \eta \\
	 & = \frac{t_k-t_0}{t_{k+1} - t_0}\  \Xii^{-1}(\bY_i,t_{k}) +\frac{1}{t_{k+1} - t_0} \frac{1}{2}\left( \bY_i + \bX_i \right) (t_{k+1}-t_k),
\end{split}
\end{align}
where $ \Xii^{-1}(\bY_i,t_{k})$ is interpolated using the neighbouring grid points. Note that \eqref{updateMeanPos} is simply a particular case of \eqref{updateLM} in conjunction with \eqref{approx2} for $\phi(\bx,t) = \bx$.
\end{itemize}

The above steps are carried out for all the grid points and iteratively in time until $t_{k+1} = t_0+T$. The completed LM is now given on scattered mean positions $\Xii^{-1}(\bX_i,t_{N})$ and can be interpolated back on the computational grid points. 

\subsection{Parameter set-up, convergence and numerical error}\label{sec:GBLA_error}

%Just like any numerical algorithm, GBLA is susceptible to error and faces some limitations. 
GBLA may be implemented on temporally and spatially decimated fields.  To distinguish between the two, we denote the (iterative) averaging timestep with $\Delta t = t_{k+1} - t_k$ and the simulation timestep (used to solve the governing equations) with $\delta t$ throughout this paper. Similarly, $\Delta \bx$ denotes the spatial resolution of GBLA, which could be coarser than the one used in simulations (i.e. weather or climate models). It is important to keep $\Delta t$ and $\Delta \bx$ smaller than the characteristic time and length scales of the fast phenomenon that is going to be filtered. To guarantee the stability of backward advection in \eqref{advect_backward} the CFL condition $ \bu \Delta t < \Delta \bx$ should be satisfied.

Just like any numerical algorithm, GBLA is susceptible to numerical errors. Denoting the local error at each iteration (from $t_k$ to $t_{k+1}$) by $e_k$, there are three main contributors to $e_k$

\begin{enumerate}
\item Finite differentiation used to advect particles backward in \eqref{advect_backward}. Considering the Taylor expansion of particle position on a trajectory, the difference between the exact and approximated particle position at time $t_k$ is $O(\Delta t^2)$.

\item Interpolation of LM at time $t_k$. In the step 3 of the algorithm, we interpolate $\Lbar{\phi}(\Xii^{-1}(\bY_i,t_k),t_k)$, leading to an error that depends on the selected interpolation scheme. Considering polynomial interpolation of degree $n$ on a grid with spacing of $\Delta x$, one can show that the interpolation error is $O((\Delta x)^{n+1})$. For example, the linear interpolation error is $O((\Delta x)^2)$ and the cubic interpolation error $O((\Delta x)^3)$. Note that the order of error is independent of spatial dimension.

\item Approximation of the last integral in \eqref{updateLM} either by \eqref{approx1} or \eqref{approx2}. The error of the former is $O(\Delta t)$ and the latter $O(\Delta t^3 (\Delta x)^{n+1})$ considering the interpolation required to calculate $\phi(\bY_i,t_{k})$.
\end{enumerate}  

Considering that there are linear leading terms in polynomial interpolation, the error in (i) directly translates into the total local error $e_k$. As a result, we can write

\begin{equation}
	e_k = O(\Delta t^2) + O((\Delta x)^{n+1}) + O(\Delta t^s),
\end{equation}
where $s =1$ or $3$ if \eqref{approx1} or \eqref{approx2} are used respectively. Hence, depending on the value of $s$, the order of local error in $\Delta t$ may be determined by (i) or (ii). The calculation of global error, which is the accumulation of local errors, is more complicated. It cannot be derived simply by multiplying $e_k$ with the number of timesteps $N = T/\Delta t$, because in the interpolation of the step 3 the neighbouring points contain the accumulated interpolation errors from the previous iterations. As a result, different sources of error become intertwined. Comparatively, the particle tracking approach does not have this type of error accumulation, because it uses the instantaneous values at neighbouring points for each timestep. Hence, GBLA might be more affected by interpolation errors than the particle tracking methods. We leave the analytical expression of global error to a future study, but in \S  \ref{sec:1Dexamples} and \S  \ref{sec:2Dflow} we calculate it numerically to investigate the convergence of GBLA.

\section{Validation}\label{sec:validationAll}

\subsection{Illustrative 1D flows}\label{sec:1Dexamples}

%\begin{figure} 
%    \centering
%    {\includegraphics[width=.6\linewidth]{StokesVel1D.eps}}
% \caption{Normalised LM of \eqref{simplestExample} calculated by GBLA over the period of $2\pi$ for different $\epsilon$.} \label{fig:first1Dexample}    
%\end{figure} 
%
%
% Considering the simple example of 
%\begin{equation}\label{simplestExample}
%u(x,t) = \epsilon \cos(x-t)
%\end{equation}
%one can derive $\Lbar{u}(x) = \Stk{u} \approx \epsilon^2/2$ for the averaging interval of $2\pi$. We compute $\Lbar{u}$ using our grid-based method and compare it with the leading-order Stokes velocity in figure \ref{fig:first1Dexample}. It is reassuring that our calculation of $\Lbar{u}$ converges to $ \epsilon^2/2$ for small $\epsilon$ as predicted by \eqref{StokesDef}. 

\begin{figure} 
    \centering
    \begin{minipage}{.5\linewidth}
         \centerline{\includegraphics[width=1\linewidth]{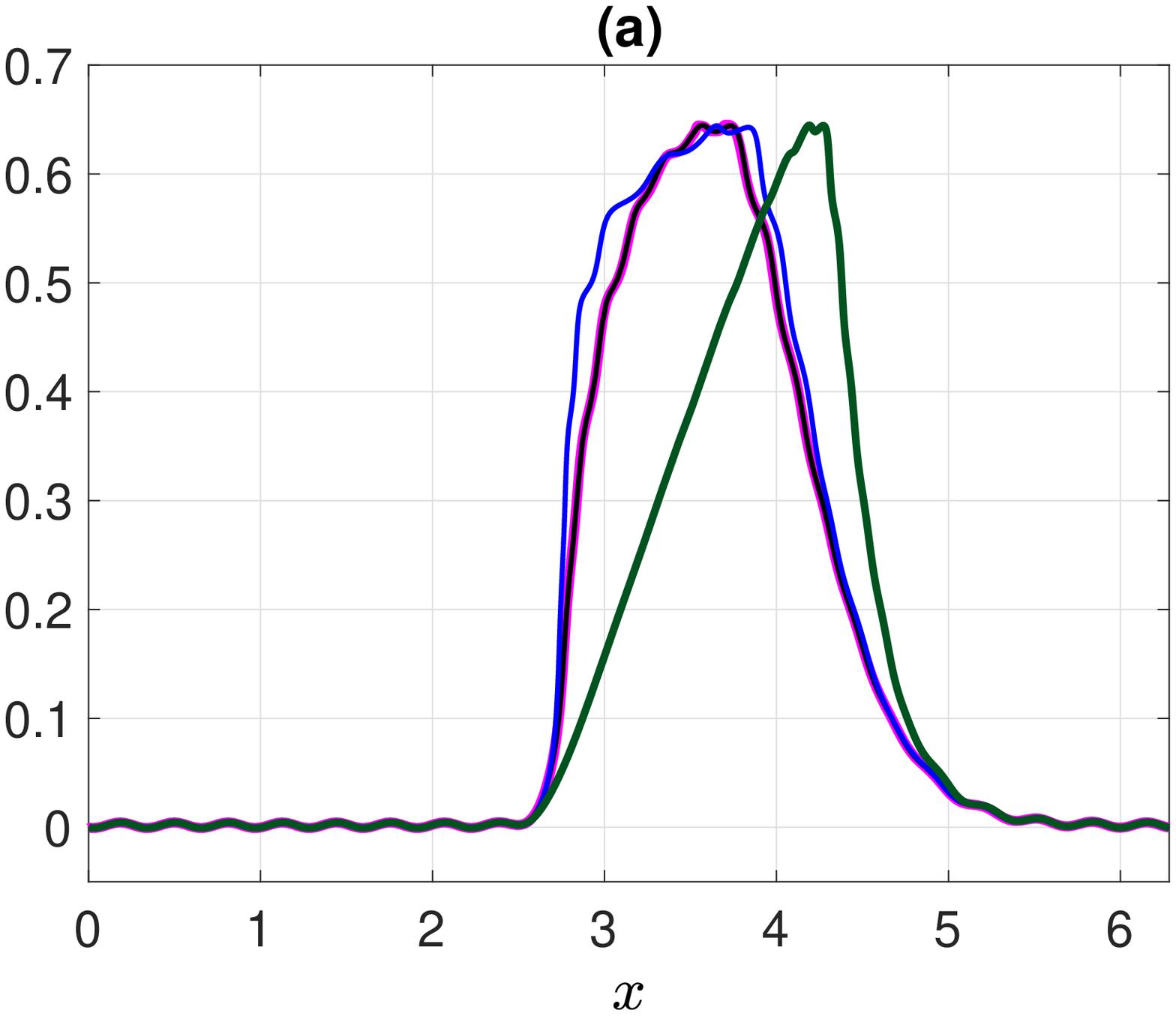}}
    \end{minipage}%
    \begin{minipage}{.5\linewidth}
        \centerline{\includegraphics[width=1\linewidth]{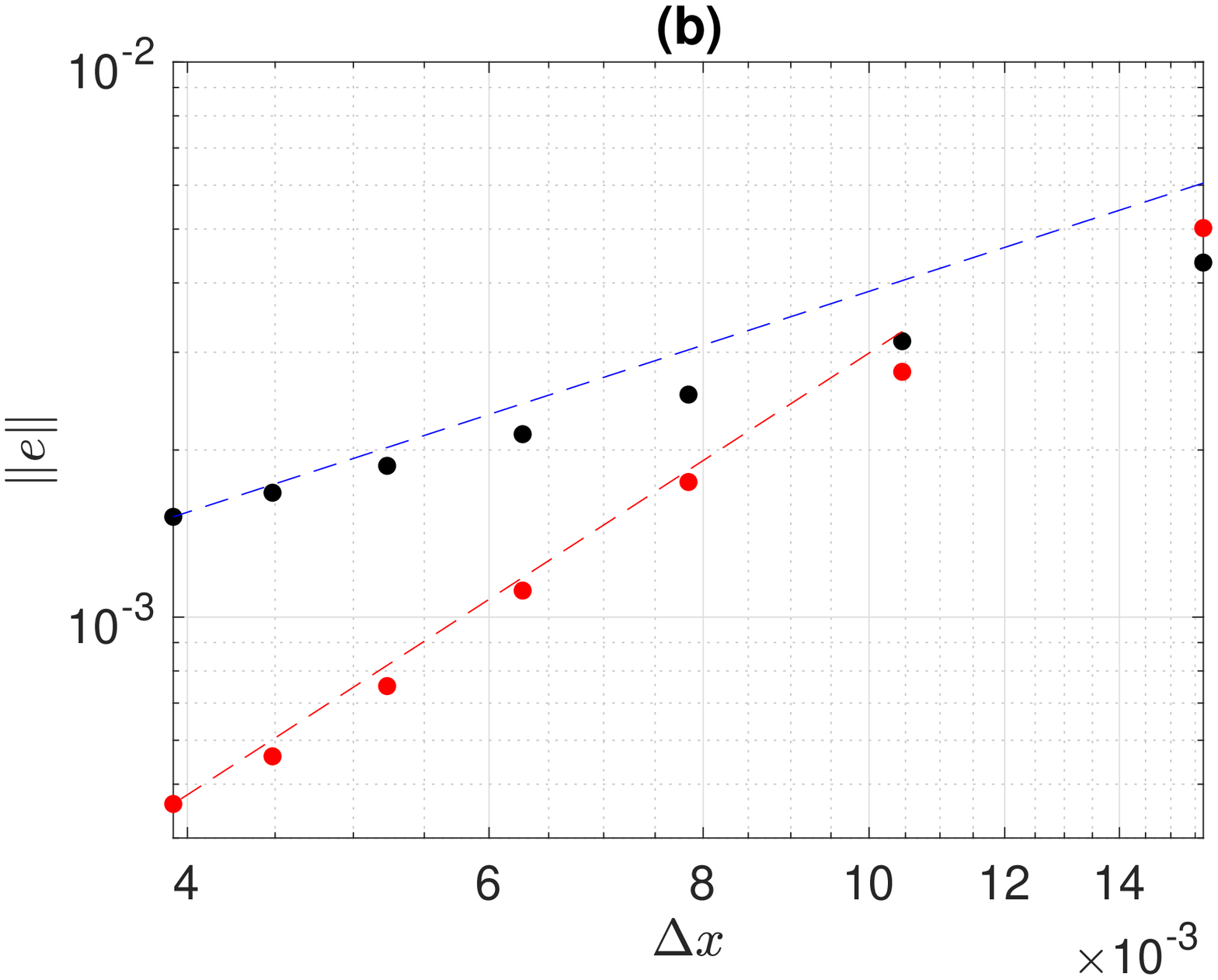}}
    \end{minipage}%    
 \caption{(a) the LM velocity of \eqref{2timescaleflow} calculated for $\epsilon = 0.06$ from $t=0$ to $t=2\pi/3$ for $\Delta x = 4.2\times 10^{-3}$ and $\Delta t = 2.6\times 10^{-3}$ using: GBLA (with cubic interpolation ) as a function of mean position (magenta), particle tracking as a function of mean position (black), particle tracking as a function of position at the middle of averaging interval (blue) and particle tracking as a function of position at the end of averaging interval (green). (b) The error norm $\Vert e \Vert$ of GBLA scheme using linear interpolation (black dots) and cubic interpolation (red dots). The slope of blue guide line is 1 and that of red line is 2.} \label{fig:second1Dexample}    
\end{figure}

We consider the following example with a velocity that consists of a fast and slow component 
\begin{equation}\label{2timescaleflow}
 u = u_{f} + u_{s}, \quad u_{f} = \epsilon \sin(20x - 20t), \quad u_{s} = \e^{-a(x-\pi)^2}, 
\end{equation}
where $a = 15$ for $x<\pi$ and  $a = 1$ for $x\geq\pi$. Figure \ref{fig:second1Dexample}(a) displays the LM  velocity of this flow regarded as a function of mean position using GBLA with cubic interpolation  (shown in magenta) and particle tracking (shown in black). For both methods, $\Delta x = 4.2\times 10^{-3}$, $\Delta t = 2.6\times 10^{-3}$ and the averaging period $T = 2\pi/3$, which is longer than the  period of the fast signal $2\pi /20$. The  GBLA and particle tracking results (magenta and black curves) perfectly lie on top of each other, which confirms the validity of GBLA. To better quantify the accuracy and convergence of GBLA, in figure \ref{fig:second1Dexample}(b) we coarsen $\Delta x$ and calculate the following error norm (representing the global error for the averaging period of $T$)

\begin{equation}\label{errorNorm_1D}
	\Vert e \Vert=\sqrt{\frac{1}{N} \sum \left(\Lbar{u}_{\text{\ GBLA}} - \Lbar{u}_{\text{\ exact}} \right)^2 },
\end{equation}
where the summation is carried out over all spatial grid point, and $N$ is the number of spatial grid points. $\Lbar{u}_{\text{\ GBLA}} $ is the LM calculated using GBLA and $\Lbar{u}_{\text{\ exact}} $ is the exact value of LM which is approximated by particle tracking at much finer spatial and temporal resolutions of $\Delta x = 2 \times 10^{-3}$ and $\Delta t = 2.6  \times 10^{-3}$. Two different interpolation schemes are used in computing GBLA for figure \ref{fig:second1Dexample}(b): linear (black dots) and cubic (red dots).  The overall message of this figure is that GBLA with different interpolation scheme converges to the exact solution if  $\Delta x$ is lowered. At smaller $\Delta x$ GBLA's error with cubic interpolation has a slope close to 2, and with linear interpolation a slope slightly shallower than 1. %This is consistent with the local errors of these schemes, which are $O((\Delta x)^3)$  and $O((\Delta x)^2)$, respectively. 
At larger $\Delta x$ other sources of error contribute to $\Vert e \Vert$ leading to shallower slopes.

\begin{figure} 
    \centering
    {\includegraphics[width=.65\linewidth]{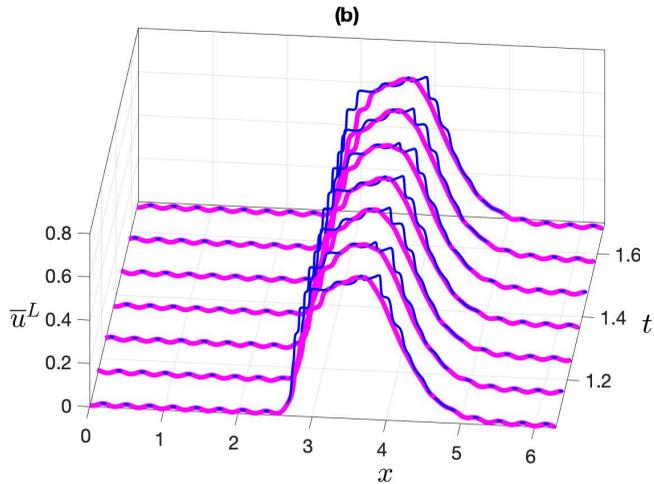}}
 \caption{(b) the LM velocity of \eqref{2timescaleflow} over the period  of $2\pi/3$, calculated for $\epsilon = 0.15$ at different times using: GBLA in terms of mean position (magenta) and particle tracking in terms of position at the middle of averaging interval (blue)} \label{fig:waterfall_1Dflow}    
\end{figure} 

As mentioned in the previous section, we choose the definition of GLM theory by requiring \eqref{GLMdisplacement} to hold and using mean position as a label for LM. We use the above example to clarify the importance of this choice. In figure \ref{fig:second1Dexample}(a), in addition to expressing the LM as a function  of mean position, we show the LM as a function of particle position at the middle of averaging interval (blue curve) and as a function of particle position at the end of averaging interval (green curve). We have used particle tracking methods to calculate these results. The black and blue curves preserve the overall shape of the slow signal, but the green curve fails to do so, which shows that the LM in terms of end position may distort the mean-flow signal. The black and blue curves have differences as well, which are more visible at the steep velocity gradient before $x=\pi$. To understand the reason behind this discrepancy, consider a particle that travels through the relatively flat part of velocity profile in the first half of the averaging interval and through the steep velocity jump in the second half. As a result, the average speed of this particle is small in the first half and large in the second half, leading to relarively small change of position in the first half compared to the second half. Therefore, the  mean position of this particle differs from its position at the middle of the averaging interval (which is closer to the position at the beginning of the interval).

In figure \ref{fig:waterfall_1Dflow} we increased the wave amplitude to $\epsilon = 0.15$ to illustrate how the GLM definition of LM can be more effective in removing the fast timescale (all other parameters are the same as those used for figure \ref{fig:second1Dexample}(a)). The LM at different times is shown in terms of the mean position calculated by GBLA (magenta curve) and in terms of the position at the middle of averaging interval calculated by particle tracking (blue curve). There is still a noticeable trace of fast signal in the blue curves, which propagates rightward in time. This is because the particle position at the middle of averaging period oscillates with the fast timescale, whereas the mean position is less affected by the these oscillations. This example is deliberately chosen to amplify the difference for various choices of spatial labels. For smaller-amplitude waves and milder background velocity gradients the magenta and blue curves converge to each other. Applying Lagrangian filtering to oceanic flows, many authors have used the final position as a label for LM \citep{shakespeare2017spontaneous,shakespeare2019momentum} or the position at the middle of averaging interval \citep{shakespeare2021LagFilt}. If a more accurate removal of fast timescale is desired, expressing LM in terms of mean position should be considered instead.

\subsection{Using materially-conserved quantity to investigate the validity of GBLA}\label{sec:validation_materiallyconserved}
In this section, we consider the materially-conserved quantities -- which are constant on each trajectory -- to study the accuracy of GBLA. For these quantities, the LM expressed in terms of final position should be equal to their instantaneous values at the end of averaging interval. We employ this fact to see how reliable GBLA is. We remind the reader that this is merely for studying the validity of GBLA; as discussed before using the GLM definition and expressing the LM in terms of mean position instead of final position is preferred when it comes to the applications of Lagrangian averaging. As a result, the step 5 of \S \ref{sec:method_description} is skipped in the calculations of this part.

\subsubsection{Conservation of vorticity in 2D incompressible flow}\label{sec:2Dflow}
\begin{figure}
     \includegraphics[width=\linewidth]{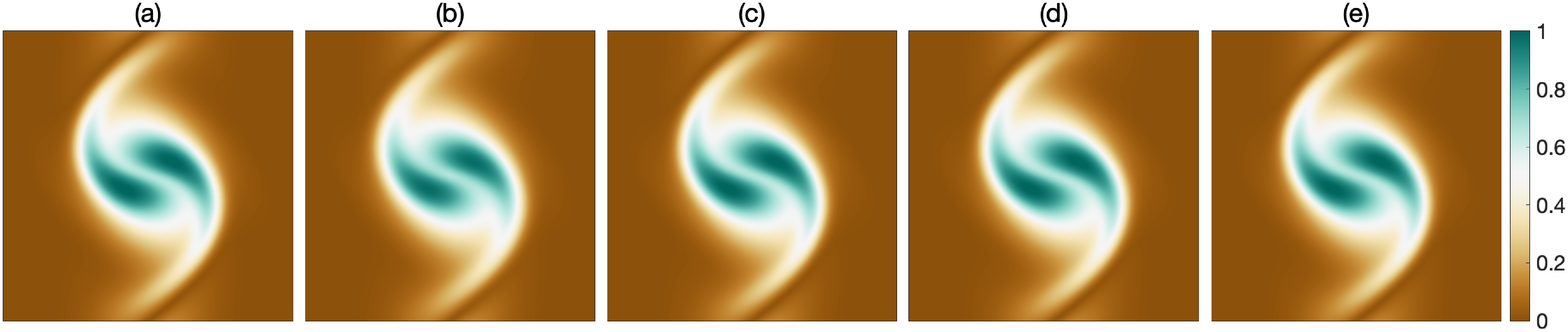}
     \includegraphics[width=\linewidth]{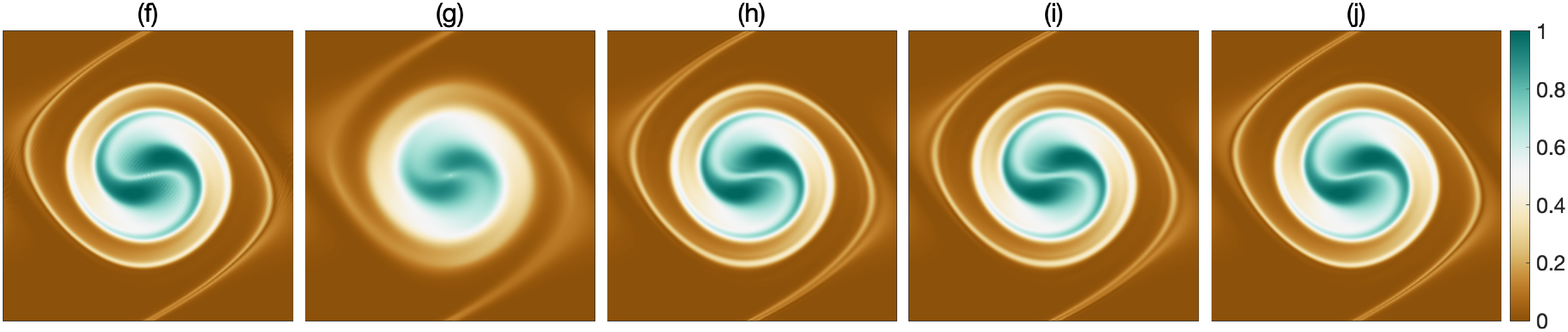}
\caption{Instantaneous and LM vorticity field for two like-signed vortices. Top row corresponds to fields at $t=10$ and bottom row to $t=30$. (a) and (f): instantaneous vorticity. (b) and (g): GBLA using linear interpolation and \eqref{approx1}. (c) and (h): GBLA using cubic interpolation and \eqref{approx1}. (d) and (i): GBLA using cubic interpolation and \eqref{approx2}. (c) and (h): LM by particle tracking and using cubic interpolation.}
\label{fig:merger}
\end{figure}

For 2D incompressible inviscid flows, vorticity is conserved on particle trajectories
\begin{equation}\label{NS2D}
 	\frac{\partial \zeta}{\partial t} + u \frac{\partial \zeta}{\partial x} + v \frac{\partial v}{\partial y} = 0, \quad \zeta = \frac{\partial v}{\partial x} -  \frac{\partial u}{\partial y}. 
\end{equation} 
%where $\bu = (u,v)$.
As a result $\zeta(\bX,t) = \Lbar{\zeta}(\Xii^{-1}(\bX,t),t)$, where $\bX = (x,y)$. Therefore, to examine GBLA we derive the LM vorticity (expressed in terms of the final particle position) and compare it with the instantaneous vorticity field at the end of averaging interval. We initialise the numerical simulation with two like-signed vortices
\begin{equation}
 	\zeta(x,y,t=0) = \e^{-(x-\pi+0.1)^2 -(y-\pi+\pi/3)^2 }  +
 	                           \e^{-(x-\pi-0.1)^2 - (y-\pi-\pi/3)^2}.  
\end{equation}
We employ a standard pseudo-spectral code to solve \eqref{NS2D} on $2\pi \times 2\pi$ domain with periodic boundary conditions using $256 \times 256$ grids points. The timestep for implementing the averaging process is 10 times coarser than the simulation timestep $\delta t = 0.002$. Figure \ref{fig:merger} portrays the LM vorticity calculated by different methods next to the instantaneous vorticity at the end of averaging interval for two different averaging periods. For the shorter averaging interval, all methods lead to very similar results. However, for the longer interval, GBLA with linear interpolation is less accurate than the others. The comparison between the panels (c)/(h) and (d)/(i) shows that the use \eqref{approx2} instead of \eqref{approx1} has not made any discernible difference, because the interpolation error exceeds other types of error due to the relatively small timestep. 

\begin{figure}
    \centering
    \begin{minipage}{.5\linewidth}
         \centerline{\includegraphics[width=1\linewidth]{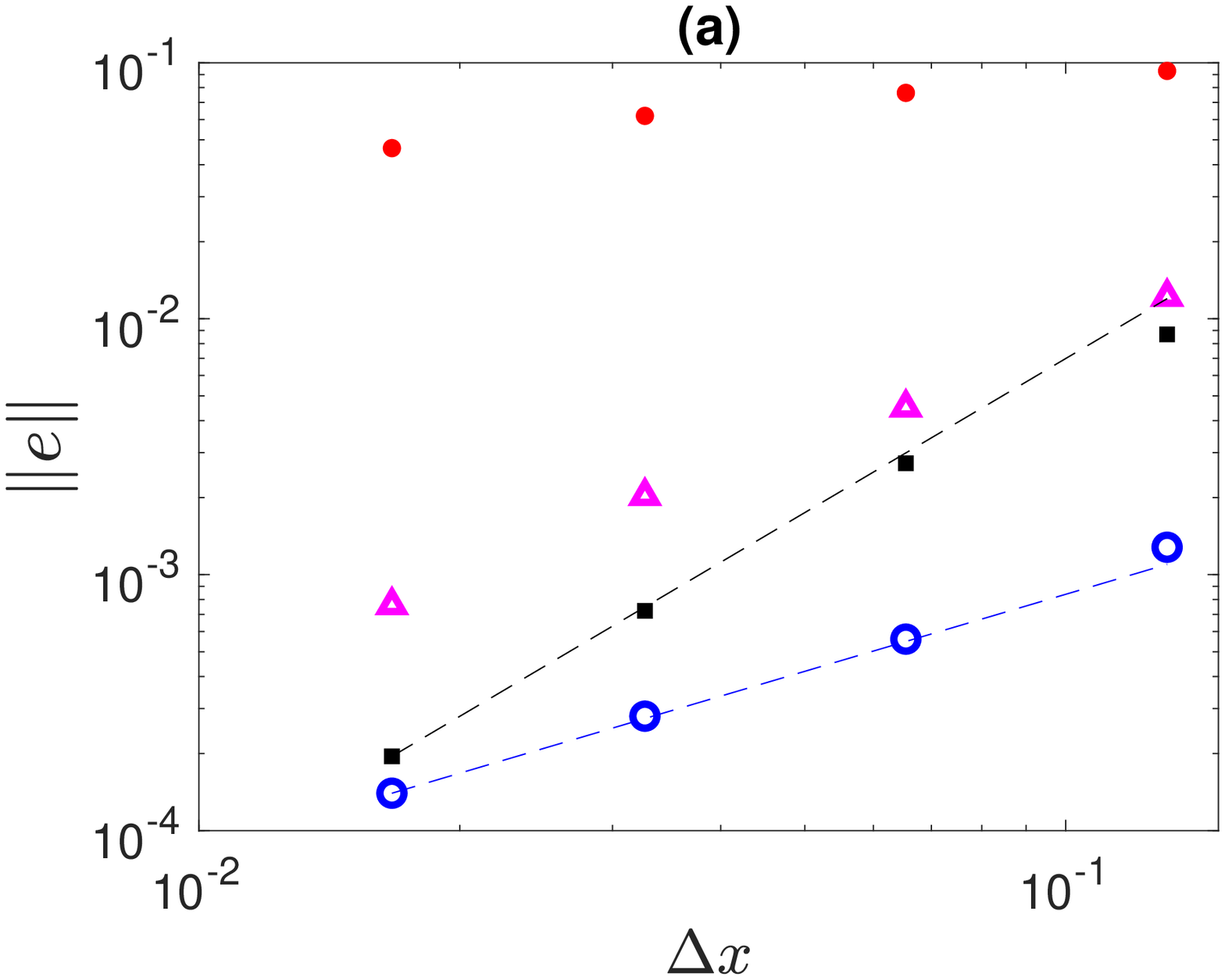}}
    \end{minipage}%
    \begin{minipage}{.5\linewidth}
        \centerline{\includegraphics[width=1\linewidth]{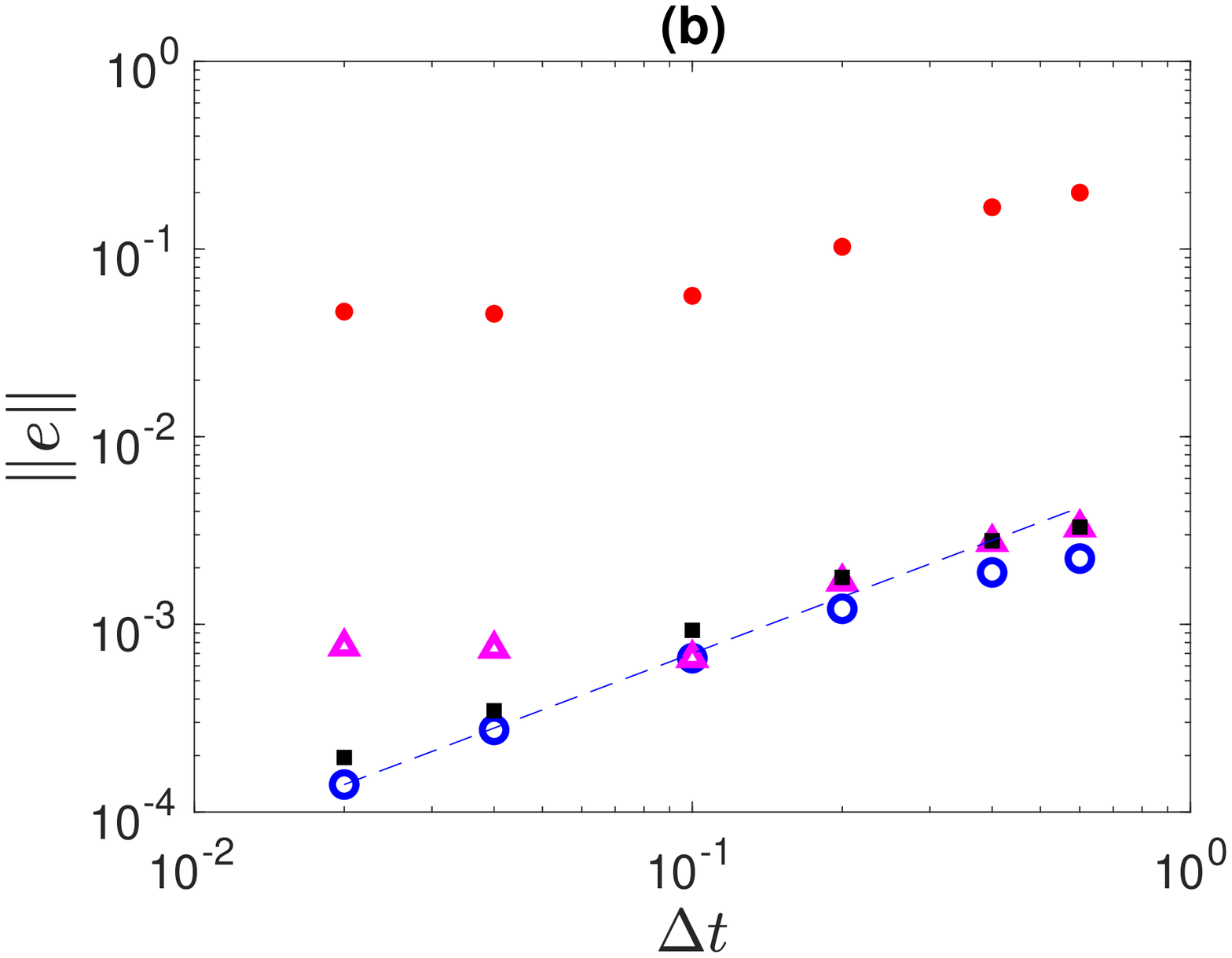}}
    \end{minipage}%    
 \caption{(a) Global error using different spatial resolution. (b) Global error using different averaging timesteps. Error norm is calculated for different schemes namely: GBLA with linear interpolation and using \eqref{approx1} (red dots), GBLA with cubic interpolation and using \eqref{approx1} (black squares), particle tracking with cubic interpolation (blue circles),  and GBLA with linear interpolation and using \eqref{approx2} (magenta triangles). The blue dashed lines have the slope of 1, and the black line has the slope of 2.} \label{fig:num_converg}    
\end{figure}

To have a more quantitative comparison between different methods, we calculate the following error norm (a 2D version of the norm used in \eqref{errorNorm_1D})

\begin{equation}
	\Vert e \Vert=\sqrt{\frac{1}{N_x N_y} \sum_i (\Lbar{\zeta}(\Xii^{-1}(\bX_i,T),T)-\zeta(\bX_i,T) )^2 },
\end{equation}
where the summation is carried out over all spatial grid point, and $N_x$ and $N_y$ are the number of grid points in each dimension. $\zeta(\bX_i,T)$ is an approximation for the exact LM, which is obtained by numerically solving \eqref{NS2D} at high resolution of $384 \times 384$ and small timestep of $\delta t = 0.002$. $\Lbar{\zeta}$ is calculated using different Lagrangian averaging methods with temporal and spatial distancing that are lowered to investigate their convergence and accuracy.

In figure \ref{fig:num_converg}(a) we set the averaging timestep to $\Delta t = 0.02 = 10\ \delta t$, which is small enough to highlight the interpolation error (see \S\ref{sec:GBLA_error} for discussion on different sources of error). GBLA with linear interpolation and using  \eqref{approx1} (red dots) underperforms GBLA and particle tracking with cubic interpolation (black squares and blue circles respectively). However, after using  \eqref{approx2} even GBLA with linear interpolation (magenta triangles) yields accurate results within the same ballpark as cubic interpolation and particle tracking approaches. With cubic interpolation, the convergence of GBLA is $O((\Delta x)^2)$, and that of particle tracking is $O(\Delta x)$. This is similar to the spatial convergence of error observed in figure \ref{fig:second1Dexample}(b) for the one dimensional synthetic flow. For large $\Delta x$ GBLA with cubic interpolation has higher errors than its particle tracking counterpart, but by lowering $\Delta x$ both methods converge to the same result.

In figure \ref{fig:num_converg}(b) the convergence of different methods with respect to $\Delta t$ is investigated. For all the calculations of this plot $\Delta x = \Delta y = 2\pi/384$. Similar to the panel  \ref{fig:num_converg}(a), the error of GBLA with linear interpolation and using  \eqref{approx1} (red dots) is almost two orders of magnitude higher than the other methods. Use of \eqref{approx2} reduces the error of linear methods by more than one order of magnitude (compare the red dots with magenta triangles).  At higher $\Delta t$, all methods seems to have a slope close to one. However, at small $\Delta t$, the methods with linear interpolation plateau as they get dominated by interpolation errors.

Considering that GBLA is an iterative algorithm, different sources of error are intertwined and tracking their effects separately is complicated. However, from the numerical investigation of global errors in figure \ref{fig:num_converg}, we can draw several insightful conclusions.  Different variations of GBLA converge robustly by lowering $\Delta x$ and $\Delta t$. GBLA is more affected by the interpolation error than the particle tracking, which can be mitigated by using cubic interpolation (or other higher order schemes). If in some applications, linear interpolations is selected to lighten the computations, the choice of \eqref{approx2} over \eqref{approx1} substantially improves the result.

\subsubsection{Conservation of PV in shallow water}\label{sec:validate_sw}

\begin{figure}
    \centering
     \includegraphics[width=.7\linewidth]{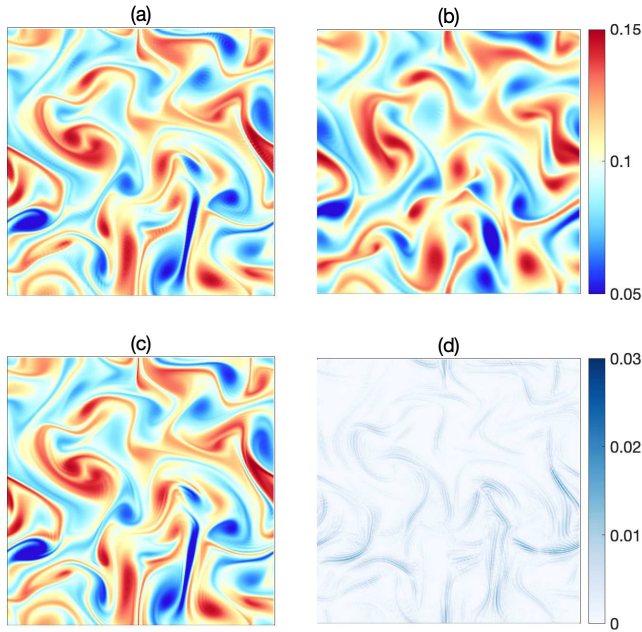}
\caption{(a) PV at the end of averaging interval $t=120$. (b) PV at $t=0$. (c) LM PV from $t=0$ to $t=120$ using GBLA with cubic interpolation scheme. (d) Absolute value of the difference between (a) and (c). Panels (a), (b) and (c) share the same colour bar, which is different from that of panel (d).}
\label{fig:pv_conserv}
\end{figure}
To put our method into a more challenging test, we consider a fully-developed turbulent flow evolving under the rotating shallow-water equations on flat topography

\begin{subequations}\label{sw_eqs}
     \begin{align}
      \frac{\partial \bu}{\partial t} + \bu \cdot \nabla \bu + \bff \times \bu &= -g  \nabla h \label{sw_momentum}\\
        \frac{\partial \bu}{\partial t}  + \nabla  \cdot (h \bu) &=  0\ , \label{sw_mass}
     \end{align}
\end{subequations}
where $h$ the height field, $g$ the gravitational acceleration and $\bff = f \hat{\boldsymbol{z}}$ the Coriolis parameter. The existence of vortices of different sizes and sharp fronts in turbulent shallow water system allow us to better examine the strengths and weaknesses of our method. For this flow, the potential vorticity (PV) 
\begin{equation}
q = \frac{\nabla \times  \bu + f}{h}
\end{equation}
is conserved on each particle trajectory. This material conservation can be used in a similar way to the previous section to assess the accuracy of Lagrangian averaging; the LM PV expressed in terms of final position should be equal to the instantaneous PV at the end of averaging interval. Selecting $f = 0.1$, $g = 0.1$ and the mean height $H = 1$, we solve the nonlinear shallow water equation using a similar pseudo-spectral method with the same computational resolution, domain size and boundary condition as in \S \ref{sec:2Dflow}. A viscous damping of the form $\nu \nabla^2$ with $\nu = 8 \times 10^{-7}$ is included for numerical stability. We initialise the simulations with a fully-developed turbulent flow which is initially in a geostrophic balance (see the initial PV in figure \ref{fig:pv_conserv}(b)). The timestep of the simulation is $\delta t = 5 \times 10^3$, and that of GBLA $\Delta t = 2 \times 10^3$. The instantaneous PV at $t=120$ in the panel (a) of figure \ref{fig:pv_conserv} is remarkably close to the LM PV over the period of $[0,\ 120]$ in the panel (c), where small scale features are also captured by GBLA (see panel (d) for the difference between the two). The small disagreement is partly caused by the viscous damping that makes the conservation of PV less accurate. This is particularly the case for the sharp fronts in panel (c) which are dampened in panel (a).

\subsection{Stokes terms for standing wave in shallow water}

\begin{figure}
    \centering
     \includegraphics[width=.95\linewidth]{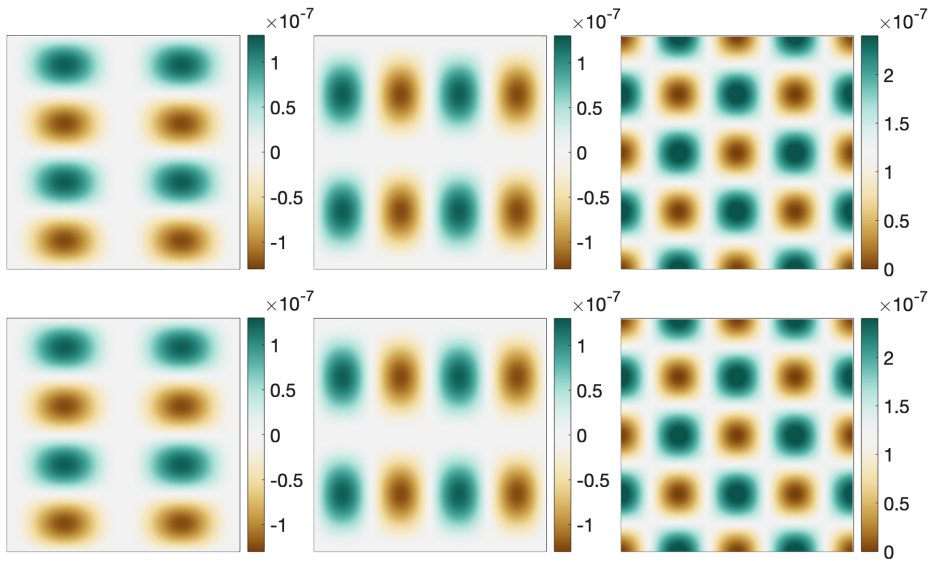}
\caption{Stokes terms computed by GBLA (top row) and from analytical expression in \eqref{Stokes_analytical} (bottom row) . Left panel $\Stk{u}$, middle panel $\Stk{v}$ and right panel $\Stk{h}$. The averaging interval is set to $2\pi/\omega$}
\label{fig:standingwave_Stokes}
\end{figure}

The difference between the Lagrangian and Eulerian mean is termed Stokes correction \citep[see e.g.][]{andrews1978exact,buhler2009waves} and can be written as 

\begin{equation}\label{StokesDef}
	\Stk{\phi} = \Lbar{\phi} - \overline{\phi} = \overline{(\xii \cdot \nabla) \phi'} + O(\epsilon^2), \quad \phi' = \phi(\bx,t) -  \overline{\phi},
\end{equation}
where the first term in the RHS is the leading order approximation for this quantity. $\epsilon$ is the amplitude of the wave term $\phi'$ after proper scaling, which is assumed $\epsilon \ll 1$. We consider a monochromatic standing wave of the form
\begin{align}
h'(x,y,t) = A \cos(\omega t) \hat{h}(x,y), \quad \hat{h}(x,y) = \cos(kx) \cos(ly), \\
\bu'(x,y,t) = \frac{g}{\omega^2 - f^2}\left( - \omega \sin(\omega t) \nabla \hat{h} + \cos(\omega t) \bff \times  \nabla \hat{h} \right), 
\end{align}
where $A$ is a small amplitude, is a solution of linear rotating shallow water equations provided that $\omega$ satisfies the dispersion relation
\begin{equation}\label{disperson_rel}
	\omega = \sqrt{f^2+g H (l^2 + k^2)}.
\end{equation} 
\cite{thomas2018wave} showed this standing wave induces a mean-flow characterised by the Stokes terms
\begin{equation}\label{Stokes_analytical}
\Stk{h} = \frac{1}{2}  \frac{g}{\omega^2 - f^2} \left| \nabla \hat{h} \right|^2, \quad 
\Stk{\bu} = -\left( \frac{g}{\omega^2 - f^2} \right)^2 \bff \times  \nabla \left( \frac{1}{2} \left| \nabla \hat{h} \right|^2 +  \frac{\omega^2 - f^2}{4 g H} \hat{h}^2 \right).
\end{equation}
The above analytical expressions provide a good test case for GBLA. We perform a numerical simulation for 
\begin{equation}
A = 0.001, \quad f=g=H=1, \quad k = l =1,
\end{equation}
and compare $\Lbar{h}-\overline{h}$ and $\Lbar{\bu}-\overline{\bu}$ calculated by the grid-based averaging method with the analytical expressions in \eqref{Stokes_analytical}. There is an excellent agreement between these two sets of quantities in figure \ref{fig:standingwave_Stokes}.

\section{Application: wave-averaged geostrophic balance}\label{sec:application}

Geostrophic balance reliably describes the slow dynamics of large-scale geophysical flows when the waves are absent or relatively small. However, in the presence of waves that are at the same order or stronger than the mean flow, this balance breaks down even after averaging over the fast timescale at fixed points. GLM theories have predicted that a modified version of geostrophic balance can still hold with coexisting strong waves if it is applied to LM quantities \citep{moore1970mass,buhler1998non-dissipative,xie2015generalised,wagner2015available,thomas2018wave}. Referring to this as wave-averaged balance, \cite{kafiabad2021wave} numerically illustrated its validity for a single barotropic vortex interacting with strong inertial waves in the context of Boussinesq equations. Instead of directly calculating the LM, they used explicit expressions of Stokes terms derived by \cite{rocha2018stimulated} in the limit of near-inertial waves. This approach circumvented the complications of particle tracking in parallelised Boussinesq simulation. However, for baroclinic flows and waves with $\omega \gg f$ explicit expressions for the Stokes terms are not available. Therefore, the LM has to be computed directly, and GBLA makes this computation straightforward. In this study, we touch upon this potential application of GBLA by investigating wave-averaged geostrophy in the context of rotating shallow water.

%\revi{
%The underlying assumption for wave-averaged geostrophy is that the flow consists of fast waves with small amplitude $\epsilon \ll 1$, interacting with a slow flow of the order $\epsilon^2$. This relative scaling of waves and slow flow is more a matter of convenience than necessity as the effect of waves on the mean-flow appears at lower orders. For waves at the same order as the slow flow similar results can be obtained by more tedious algebra and expansions to higher orders. Using this assumption, in appendix we show
%\begin{equation}\label{waver_balance}
% \bff \times \Lbar{\bu} = -g \Lbar{\nabla h}.
%\end{equation}
%which is a form of wave-averaged geostrophy that is slightly different than the existing expressions in the literature. We emphasise that $\Lbar{\nabla h}$ is the LM of the height gradient which is different from the gradient of the LM height $\nabla \Lbar{h}$. The RHS of \eqref{waver_balance} can also be written as $ -g \nabla ( \overline{h} + \Stk{h}/2)$  (see e.g. \cite{thomas2018wave} or \cite{wagner2015available} for the counterpart pressure gradient in the Boussinesq equations). We prefer to keep the current form of \eqref{waver_balance} as expressions in terms of $\Stk{h}/2$ unnecessarily complicates the computation and derivation.
%}

We first present a brief derivation of wave-averaged balance for the momentum equations in a form that is slightly different than the existing expressions in the literature (a more detailed derivation for shallow water can be found in \cite{thomas2018wave} and for the Boussinesq equations in \cite{wagner2015available}). We then quantify to what extent this balance holds for waves with different parameters and compare it with Eulerian-averaged geostrophic balance. Similar to preceding studies, the underlying assumption in our derivation is that the flow consists of fast waves with small amplitude $\epsilon \ll 1$, interacting with a slow flow of the order $\epsilon^2$. This relative scaling of waves and slow flow is more a matter of convenience than necessity as the effect of waves on the mean-flow appears at lower orders. For waves at the same order as the slow flow similar results can be obtained by more tedious algebra and expansions to higher orders. Considering these assumptions, we expand flow variables as
\begin{equation}
\bu = \epsilon \bu _1 + \epsilon^2 \bu _2 + O(\epsilon^{3}),
\end{equation}
where  $\bu _1 $ denotes the wave terms with $\overline{\bu _1} = 0$, and the mean-flow at the leading order is $\overline{\bu } = \epsilon^2 \overline{\bu _2} + O(\epsilon^{3})$ (similarly expansion is carried out for $h$). Additionally, the mean-flow is assumed to vary on the  slow timescale of quasigeostrophic dynamics, leading to $\partial  \overline{\bu _2}/ \partial t = O(\epsilon^2)$. At order $\epsilon$, the expansion of \eqref{sw_momentum} leads to
\begin{equation}\label{ord1moment}
\frac{\partial \bu _1}{\partial t} + \bff \times  \bu _1 =  -g  \nabla h_1 .
\end{equation}
%which are simply the momentum equations for linear shallow-water waves. 
The next-order equation after averaging is
\begin{equation}\label{ord2moment}
 \overline{\bu _1 \cdot \nabla \bu _1} + \bff \times  \overline{\bu _2} = -g \overline{\nabla h_2}.
\end{equation}

After taking the time derivative of \eqref{mapping} and some algebraic manipulation, one can show that $\bu _1 = \partial \xii / \partial t +  O(\epsilon)$ (see \cite{andrews1978exact} for the detailed derivation). As a result,
\begin{equation}\label{nonlilnear_ord2a}
\overline{\bu _1 \cdot \nabla \bu _1} =   - \overline{\xii \cdot \nabla \frac{\partial \bu _1}{\partial t}}.
\end{equation}
Substituting $ \partial \bu _1  / \partial t $ from \eqref{ord1moment} in the above equation and using \eqref{StokesDef}, we derive
\begin{equation} \label{nonlilnear_ord2}
	\overline{\bu _1 \cdot \nabla \bu _1} =  \bff \times  \overline{\xii \cdot \nabla \bu _1}  + g \overline{\xii \cdot \nabla (\nabla h_1)} = \bff \times \Stk{\bu} + g \Stk{\nabla  h},
\end{equation}

 %Note that we consider the Stokes correction of the height gradient $\Stk{\nabla  h}$ which is different than the gradient of Stokes height $\nabla \Stk{h}$. 
\eqref{nonlilnear_ord2} and \eqref{ord2moment} lead to
\begin{equation}\label{pre_final}
 \bff \times \left(  \overline{\bu} + \Stk{\bu} \right)= -g \left(  \overline{\nabla h} + \Stk{\nabla h} \right),
\end{equation}
where we replaced $\overline{\bu _2}$ with $\overline{\bu}$ (and removed the book-keeping parameter $\epsilon$). Rewriting \eqref{pre_final} in terms of LM variables gives to the final form of wave-averaged geostrophy
\begin{equation}\label{waver_balance}
 \bff \times \Lbar{\bu} = -g \Lbar{\nabla h}.
\end{equation}
We emphasise that $\Lbar{\nabla h}$ is the LM of the height gradient which is different from the gradient of the LM height $\nabla \Lbar{h}$. The RHS of \eqref{waver_balance} can also be written as $ -g \nabla ( \overline{h} + \Stk{h}/2)$  (see e.g. \cite{thomas2018wave} or \cite{wagner2015available} for the counterpart pressure gradient in the Boussinesq equations). We prefer to keep the current form of \eqref{waver_balance} as expressions in terms of $\Stk{h}/2$ unnecessarily complicates the computation and derivation.

To verify \eqref{waver_balance}, we study two examples of slow flow interacting with strong waves. In the first example, we consider an initial condition that consists of a Gaussian anticyclone similar to what studied by \cite{kafiabad2021wave}. The vertical vorticity of this vortex is defined as
\begin{equation}\label{vort_profile}
\zeta(r, t = 0) = - \zeta_{\mathrm{max}} \, \e^{-r^2/a^2}, 
\end{equation}
where $r$ is the radial coordinate, $a$ the vortex radius, and $\zeta_{\mathrm{max}}$ the initial maximum vorticity. The velocity fields associated with this vorticity is set to zero far from the vortex, and the associated height field is defined such that the geostrophic balance is initially maintained. This vortex is superimposed with a Poincar\'e wave of the form
\begin{equation}\label{initial_wave}
 u' =  A \cos x, \quad
 v' =  \frac{f}{\omega} A \sin x, \quad
 h' = H + \frac{H}{\omega}   A \cos x,
\end{equation}
where $\omega$ satisfies \eqref{disperson_rel} (for $k=1$ and $l = 0$). Selecting $a = 0.4$, $f= 20$, $g = H = 1$, $\zeta_{\mathrm{max}} = 0.5$ and $A = 3$, we numerically solve the nonlinear shallow-water equations for three periods of the initial wave and time-average the variables over this interval ($3 \times 2\pi/\omega$). The time step -- which is the same for both the simulation and averaging -- is set to $10^{-4}$, and all other parameters are kept the same as those in \S \ref{sec:validate_sw}. 

\begin{figure}
\centering
\includegraphics[width=\linewidth]{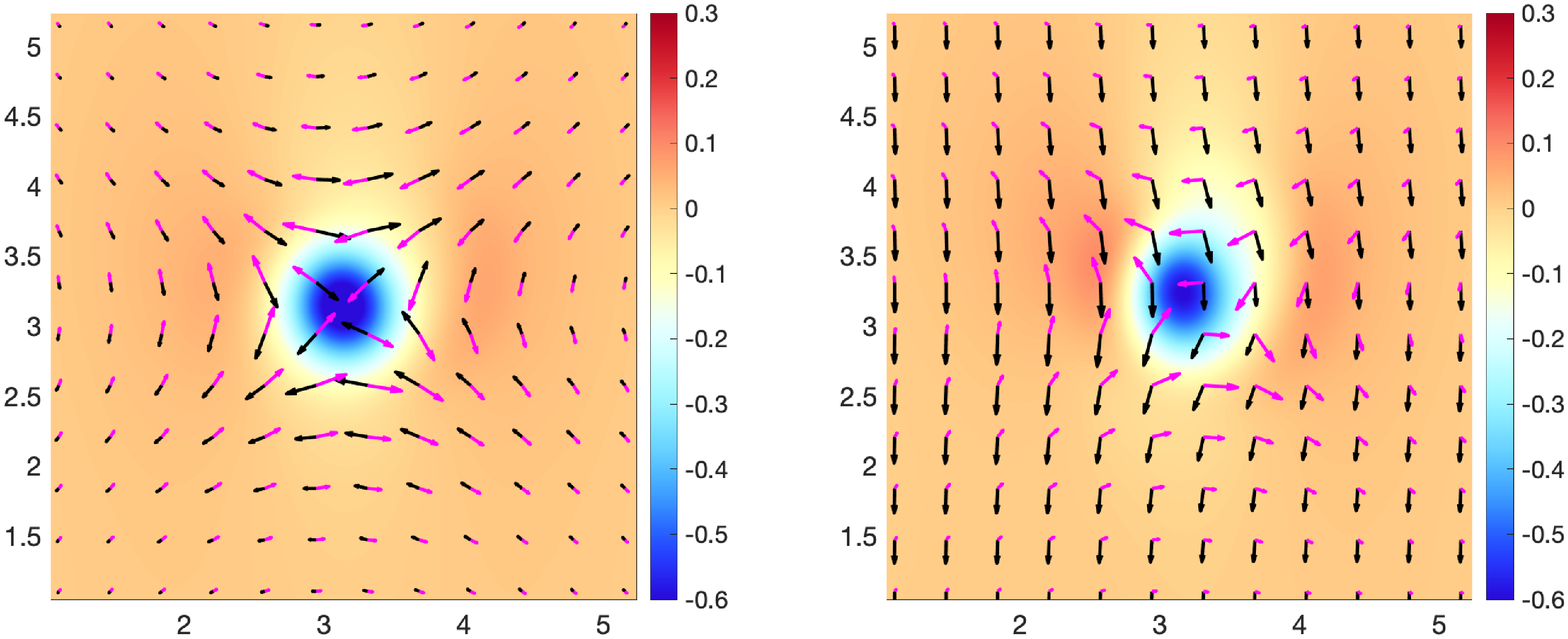}
\caption{Black arrows:  $\bff \times \Lbar{\bu}$ (left) and $\bff \times \overline{\bu}$ (right). Magenta arrows: $g \Lbar{\nabla h}$ (left) and $g {\nabla \overline{h}}$ (right). Background colour:  $\nabla \times \Lbar{\bu}$ (left) and $\nabla \times \overline{\bu}$ (right). Only a part of resolved domain is shown. }
\label{fig:quiverplot}
\end{figure}

Figure  \ref{fig:quiverplot} portrays the vectors of Lagrangian- and Eulerian-mean Coriolis forces and height gradients (scaled by $g$) using GBLA. It is clear that the geostrophic balance holds very well for the LM values as described by \eqref{waver_balance}, whereas the Eulerian-mean version has broken down. In the background $\nabla \times \Lbar{\bu}$ and $\nabla \times \overline{\bu}$  are shown, which are slightly different fields. Unlike the initial vortex, $\nabla \times \overline{\bu}$ (and to a lesser extent $\nabla \times \Lbar{\bu}$) are not completely radially symmetric due to the wave feedback on the mean-flow.

\begin{figure}
\centering
\includegraphics[width=\linewidth]{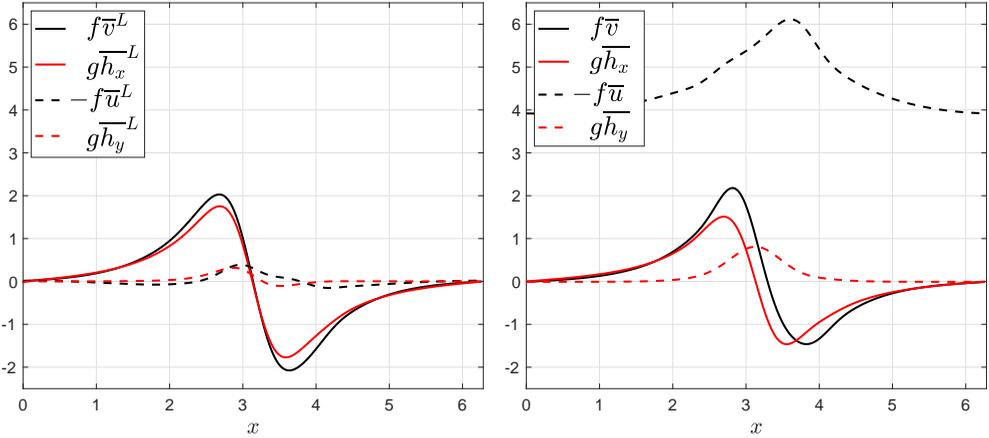}
\caption{The profiles of different components of Coriolis force and (scaled) height gradient at $y=0$. All the quantities in the left are Lagrangian averaged, and in the right Eulerian averaged.}
\label{fig:cross-section}
\end{figure}

To better investigate the accuracy of \eqref{waver_balance}, we take a cross-section of this flow at $y=0$ in figure \ref{fig:cross-section} and show the profiles of Coriolis force and (scaled) height gradient. The most outstanding difference in this figure is between $-f \overline{u}$ and $g \overline{h_y}$ (the dashed lines in the right panel), which is expected from the form of the initial wave. Let us assume a hypothetical scenario that the linear wave in \eqref{initial_wave} does not interact with the vortex. The following Stokes expressions can be derived for this linear wave
\begin{subequations}\label{waves_Stokes}
     \begin{gather}
 		f\ \overline{ \xii \cdot \nabla  u'} =  f\ \overline{\frac{-A}{\omega} \sin (x-\omega t) \frac{\partial}{\partial x} A\cos(x-\omega t)} = f \frac{A^2}{2 \omega}  \approx 4.5, \\
 		f\ \overline{ \xii \cdot \nabla  v'} = g\ \overline{ \xii \cdot \nabla h_x} = g\ \overline{ \xii \cdot \nabla  h_y} = 0,
 	\end{gather}
\end{subequations}
where the averaging is carried over one or several wave periods. In the presence of vortex, the fast component of the flow changes in time leading to more complicated Stokes expressions than those in \eqref{waves_Stokes}. Nevertheless, these values can provide a ballpark estimate for the difference between the Eulerian and Lagrangian velocities. This explains the significant difference between $-f \overline{u}$ and $g \overline{h_y}$ as we expect $\Stk{u}$ to be the largest Stokes term from \eqref{waves_Stokes}. In fact, several other numerical simulations with different wave amplitudes have been performed to verify that this difference proportionally changes with $ f A^2 / (2 \omega)$. Leaving this major difference aside, $f \Lbar{v}$ and $g \Lbar{h_x}$ are also closer to each other than their Eulerian counterparts. Even after removing the constant shift between $-f \overline{u}$ and $g \overline{h_y}$, $- f \Lbar{u}$ more closely follows $g \Lbar{h_y}$. These observations illustrate the merits of wave-averaged balance over classical geostrophic balance when strong waves are present.

\begin{figure}
\centering
\includegraphics[width=\linewidth]{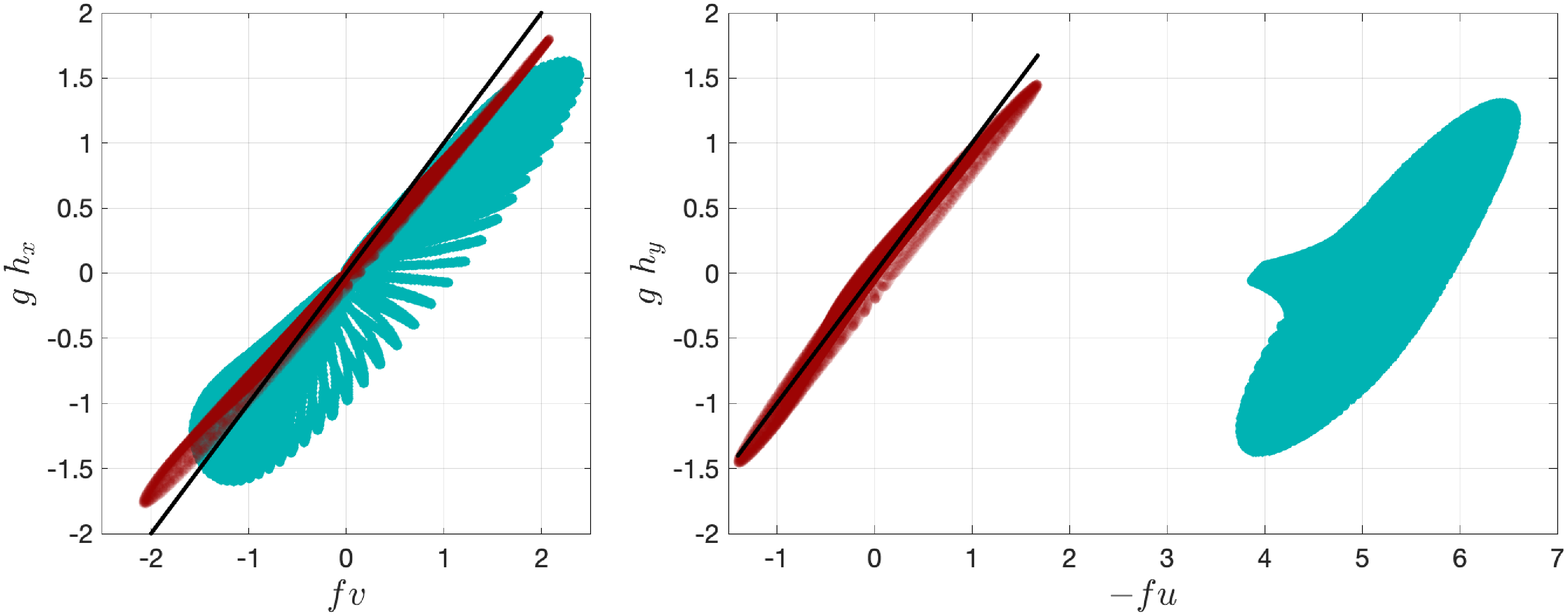} \\
\includegraphics[width=\linewidth]{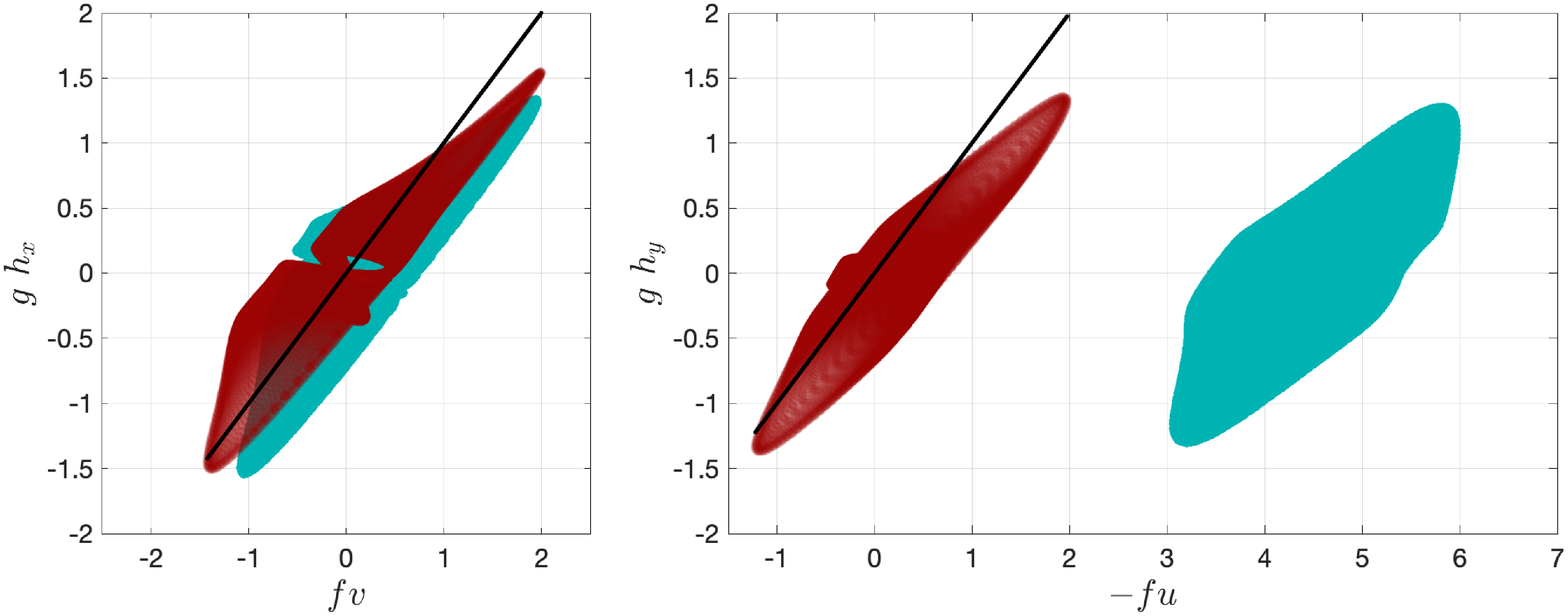} \\
\includegraphics[width=\linewidth]{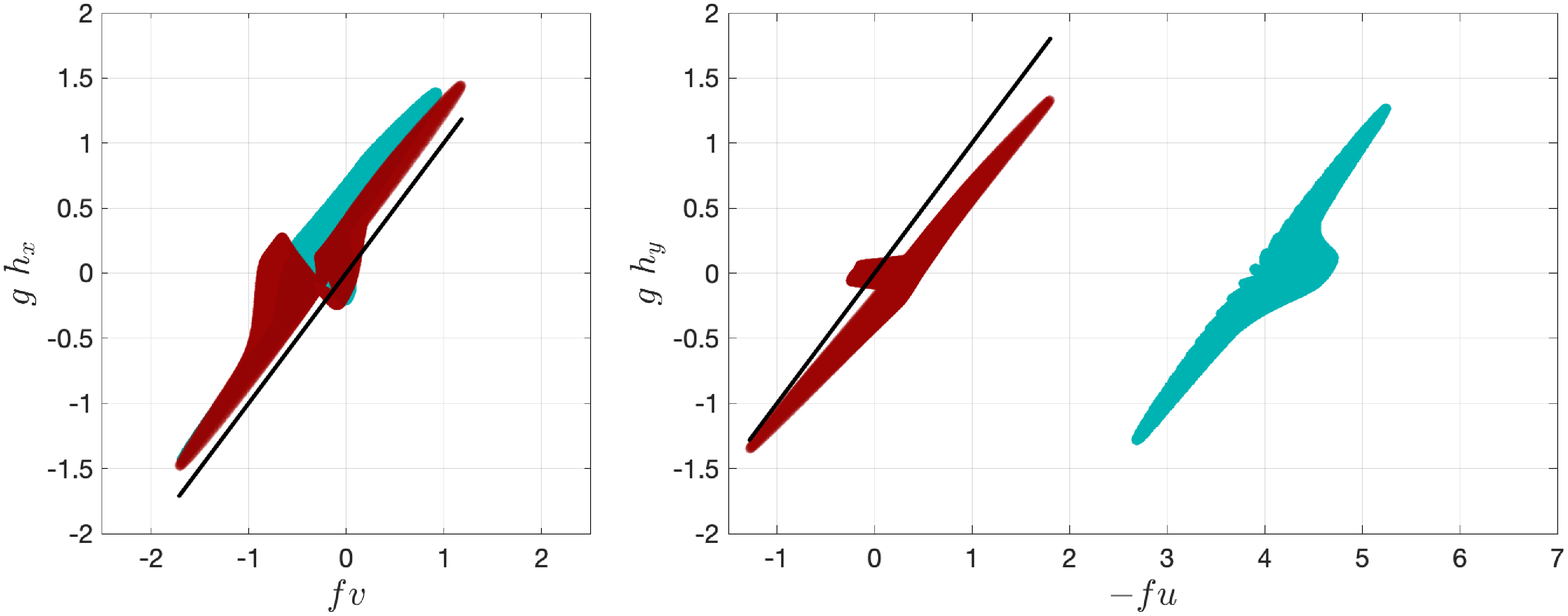}
\caption{The scaled height gradient as a function of the Coriolis force after Lagrangian averaging (in red) and Eulerian averaging (in green). The black line is $y=x$, which represents the geostrophic balance. Every marker represents the mean values for a computation grid point in space. Top row: $g=H=1$, middle row: $g=H=6$, and the bottom row: $g=H=10$. All other parameters are kept the same as those in figure \ref{fig:quiverplot} and \ref{fig:cross-section}.}
\label{fig:geospread}
\end{figure}

Figure \ref{fig:geospread} displays the scatter plots of Lagrangian- and Eulerian-means for $g h_x$ and $g h_y$ as functions of $f v$ and $-f u$ respectively (using all the computational grid points). For perfect geostrophic balance, one expects the data points to lie on $y=x$ (shown in black in figure  \ref{fig:geospread}). We consider three simulations with $g=H=1$, $g=H=6$ and $g=H=10$ leading to wave frequencies of $\omega = 20.03$, $20.88$ and $22.36$ respectively. It is clear that the geostrophic balance holds much better for the LM quantities as the red markers are closer to $y=x$ than the green ones for all three cases. Considering these different wave frequencies implies that \eqref{waver_balance} is a good approximation in the presence of non-inertial waves as well as inertial waves. Unfortunately, we could not investigate much higher wave frequencies due to the formation of shocks at the high wave-amplitude, which is considered in our simulations.

\begin{figure}
\centering
\includegraphics[width=\linewidth]{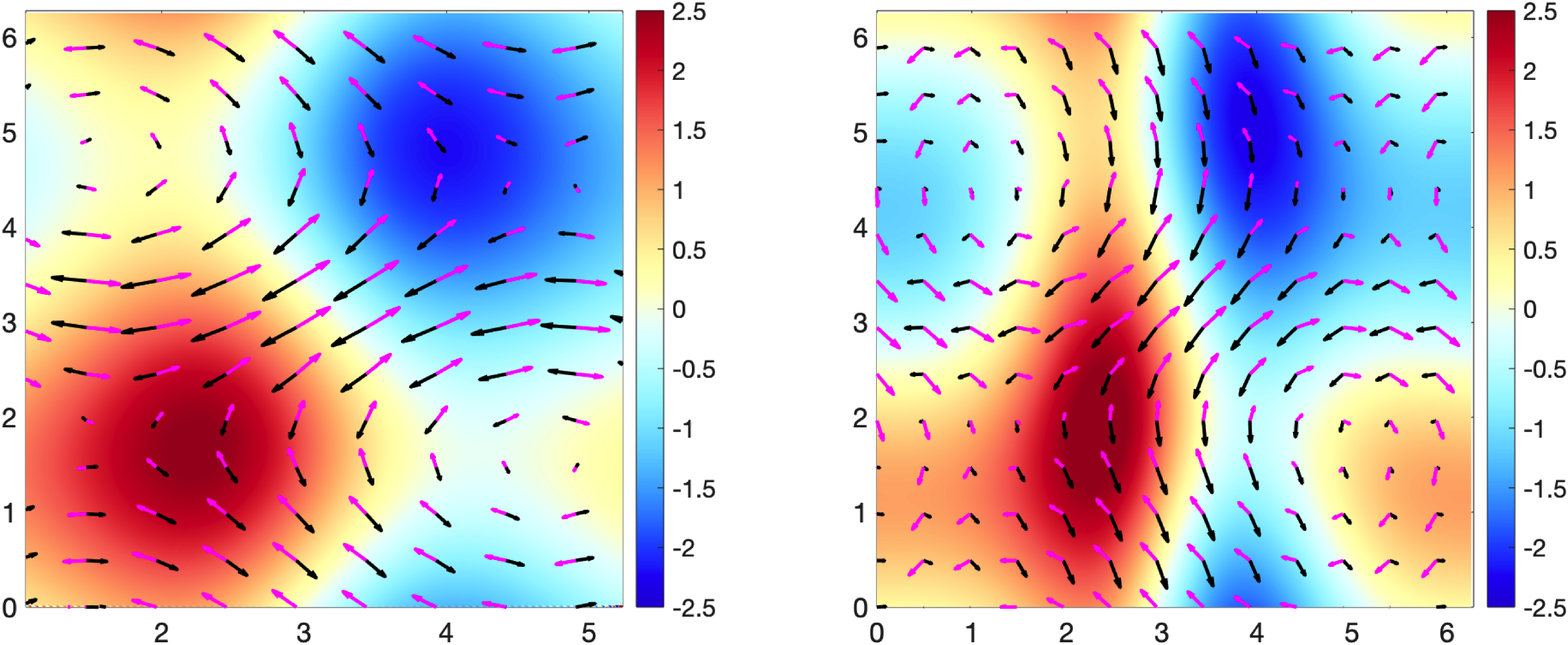}
\caption{Similar figure to \ref{fig:quiverplot} for two opposite-sign vortices described in \eqref{2vortices}. The entire computational domain is shown in this figure.}
\label{fig:quiverplot_dipole}
\end{figure}

\begin{figure}
\centering
\includegraphics[width=\linewidth]{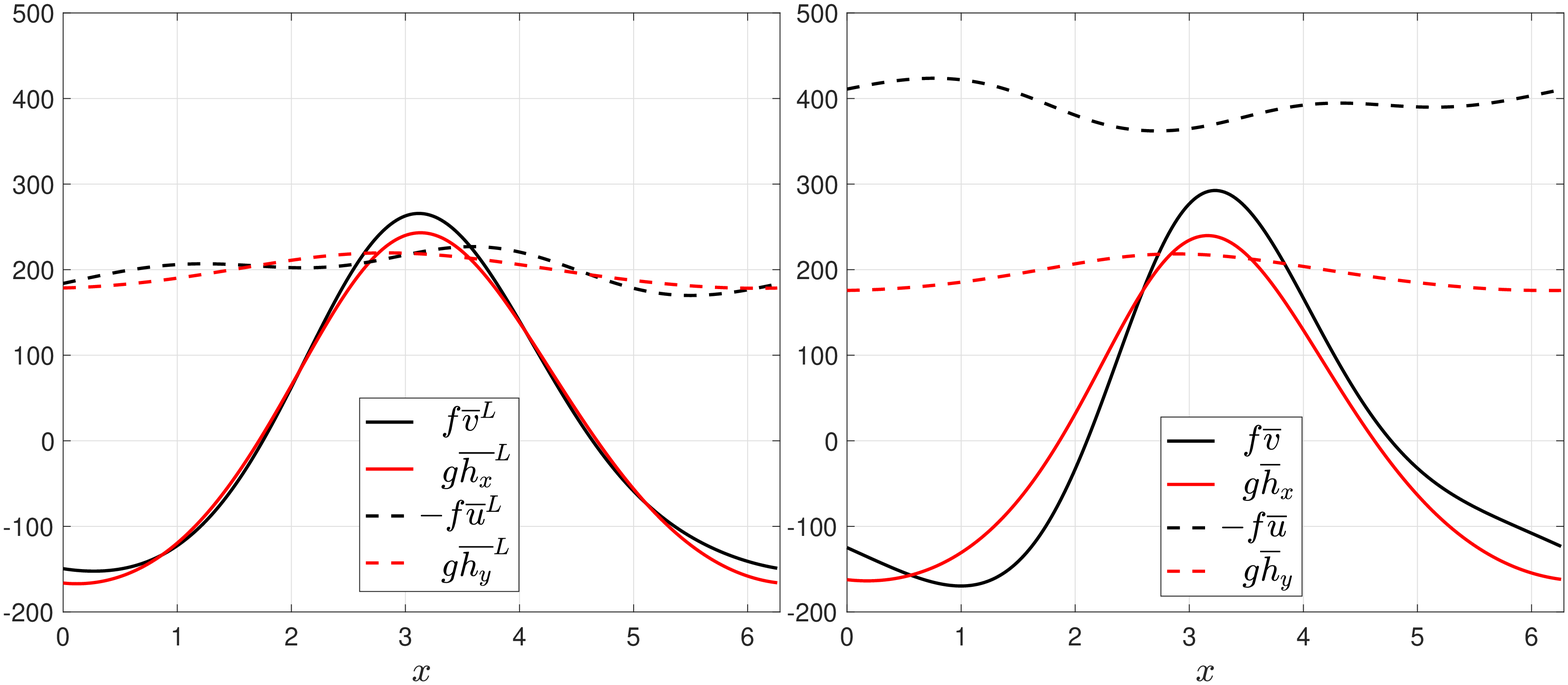}
\caption{Similar figure to \ref{fig:cross-section} for two opposite-sign vortices described in \eqref{2vortices}.}
\label{fig:cross-section_dipole}
\end{figure}

\begin{figure}
\centering
\includegraphics[width=\linewidth]{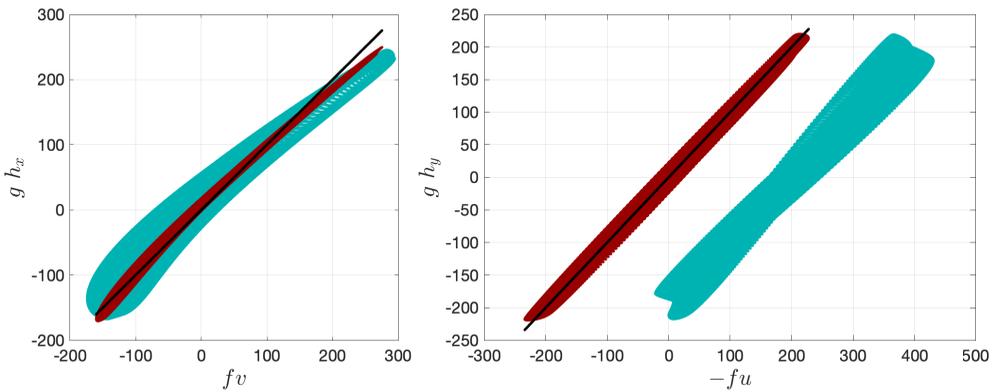} 
\caption{Similar figure to \ref{fig:geospread} for two opposite-sign vortices described in \eqref{2vortices}.}
\label{fig:geospread_dipole}
\end{figure}

In the next example, we initialise the slow flow with two opposite-sign vortices
\begin{equation}\label{2vortices}
u = -\cos y, \quad v =  \cos x , \quad h =  - \frac{f}{g} (\sin x+\sin y).
\end{equation}
A similar form of wave to \eqref{initial_wave} is superimposed with these vortices. We set the flow and wave parameters as $f= 100$, $g = 100$, $H =4$ and $A = 20$. All other parameters and numerical set-up is kept the same as the previous example. In figure \ref{fig:quiverplot_dipole}, we observe that the vectors of $\bff \times \Lbar{\bu}$ and $g \Lbar{\nabla h}$ balance each other very well (left panel), whereas the balance for their Eulerian counterparts breaks down (right panel). The background fields indicate that $\nabla \times \overline{\bu}$ is more affected by the wave signal than $\nabla \times \Lbar{\bu}$ is. Further assurance comes from the figure \ref{fig:cross-section_dipole} where the LM Coriolis force components much more closely follow the LM scaled height-gradient components. Figure \ref{fig:geospread_dipole} also verifies that the Eulerian-mean quantities are farther from $y=x$ and wider-spread.

%Setting
%\begin{equation}
%a = 0.001, \quad f=g=H=1, \quad k = l =1,
%\end{equation}
%we initialise the shallow water equations with 
%\begin{equation}
%h(x,y,t=0) =  a \cos(x) \cos(y), \quad \bu(x,y,t=0) = \frac{g}{\omega^2 - f^2} \bff \times  \nabla(\cos(x) \cos(y))
%\end{equation}
%For this set-up, the Stokes terms simplify to 
%\begin{equation}
%\Stk{h} = \frac{1}{4} \left( \sin^2 x \ \cos^2 y + \cos^2 x \ \sin^2 y \right), \quad \Stk{u} = -\frac{a^2}{8} \sin 2y \sin^2 x, \quad \Stk{v} = \frac{a^2}{8} \sin 2x \sin^2 y
%\end{equation}
%
%In figure \ref{fig:standingwave_Stokes}, we compared these analytical expression with the direct calculation of $\Lbar{h}-\overline{h}$ and $\Lbar{\bu}-\overline{\bu}$ using our grid-based averaging method. There is an excellent agreement for $\Stk{h}$ and a less striking agreement for Stokes velocities.

\section{Discussion}\label{sec:discussion}

Through several examples, we find that GBLA reliably calculates the LM of flow variables and has several advantages over its particle tracking counterparts. Particle-tracking method are perform either forward or backward in time to calculate LM fields. The forward-time tracking cannot guarantee that the particles at a later time are uniformly distributed. In fact, in regions of confluence the particles get cluttered in a small part of domain. The backward particle tracking does not suffer from this issue but requires storing times series of several fields: the instantaneous value of the desired field, its interpolated values at particle positions, particle positions along the trajectories and velocity components. GBLA does not need to store any time series; it only stores a few fields such as partial means at two timesteps. As a result, it can significantly  reduce the memory consumption if a large number of grid points or timesteps are employed.  For the LM of any field other than velocity, GBLA does not require interpolation of velocities, whereas the particle tracking methods do. For instance, assume that the LM temperature needs to be calculated. For particle tracking, in addition to the temperature, the velocity field needs to be interpolated at particles instantaneous positions to advect them backward or forward. However, for GBLA there is no need for interpolating velocities, and only one set of interpolation for temperature partial means is needed. As a result, the amount of interpolation in GBLA is at least half of those in particle tracking. Depending on the spatial dimension, it could be even less than half. For example in three dimensional flows, it is only a quarter of interpolations required by particle tracking.  

All the advantages said, we caution that the GBLA results could be affected by interpolation errors for long averaging periods or coarse spatial resolutions; better interpolation schemes and finer timesteps can significantly reduce these errors. Our numerical investigations establishes a robust spatial and temporal convergence of GBLA even for linear interpolation. Hence, by using finer grids one obtains reliable LM fields.   %Another advantage of GBLA is maintaining the load balance and reducing the communications between processes in parallelised solvers, which was discussed in \S \ref{sec:intro}. 

In this study, we explained the averaging process only for a single interval, but in some applications the evolution of mean-flow or wave field in time is desired. This can be achieved by a moving time-window that frees up the memory from completed intervals and allocates it to the new intervals. Note that GBLA does not require the discretisation in time or space to be uniform. Hence, it can be improved by adaptive time stepping or more advanced interpolation scheme. Another direction of improvement is using more advanced filtering schemes. Many studies in geophysical fluids employ low-pass filtering in Fourier space \citep[see e.g.][]{shakespeare2017spontaneous,shakespeare2021LagFilt}, which can be more effective than simple averaging used in \eqref{LagMean}. This filtering (as well as other advanced filtering schemes) can be achieved by using GBLA, if an appropriate kernel is included in the definition of mean in \eqref{LagMean}. Considering other definitions of mean/perturbation than the one used in \eqref{LagMean} and  \eqref{GLMdisplacement} \citep{soward2010hybrid,gilbert2018geometric}, it is an open question to the author whether a similar approach to GBLA can be used to derive these forms of mean.

We mainly focused on the description and validation of our method and touched upon one application of GBLA. We leave other applications of GBLA for future work. Recent studies such as \cite{shakespeare2017spontaneous,shakespeare2018life,shakespeare2019momentum} have demonstrated that Lagrangian filtering is more effective in extracting wave signals from the flow. GBLA can make Lagrangian filtering more accessible by making the parallel implementation easier and reducing the memory footprint. Another application of GBLA is the study of the reduced models described for LM quantities \citep[e.g.][]{xie2015generalised,wagner2015available}, which was briefly considered in this paper. Studying these models in the context of Boussinesq equations is often challenging due to complications of particle tracking in high-resolution parallelised solvers. GBLA can also be applied to observations or more realistic Global Circulation Models (GCM) to gain more insight about the ocean and climate dynamics.

Ultimately, GBLA can improve the parameterisation schemes used in ocean and weather models that need to dissipate energy at much larger scales than viscous scales. Such dissipation -- which is required for numerical stability and closing the energy budget-- reduces the effective resolution of these models \citep[see e.g.][]{skamarock2004evaluating}. Several studies have shown that the unbalanced and fast processes such as waves dynamics initiate the downscale flux of energy toward dissipation range in the ocean and atmosphere \citep{waite2009mesoscale,barkan2017stimulated,kafiabad2018spontaneous,taylor2020effects,xie2020downscale}. Hence, one can argue that the dissipation of the fast component after removing the LM preserves the slow dynamics better than the less selective dissipation of all fields or their Eulerian high-pass part.

\backsection[Acknowledgements]{Jacques Vanneste is thanked for insightful discussions and suggestions that greatly improved the manuscript.}

\backsection[Declaration of Interests]{The author reports no conflict of interest.}

\bibliographystyle{jfm}
\bibliography{myGFD.bib}

\end{document}